\documentclass[a4paper]{article}

\usepackage{amsmath,amsfonts}
\usepackage{natbib}
\usepackage{graphicx}

\newcommand{\ind}{{\text{\texttt{I}}}}
\DeclareMathOperator{\Prob}{pr}
\DeclareMathOperator{\Expec}{E}
\DeclareMathOperator{\Var}{Var}

\newcommand{\unitmatr}{{\rm Id}}

\newcommand{\dint}{{\textup d}}
\newcommand{\YI}{Y_{I}}
\newcommand{\YJ}{Y_{J}}
\newcommand{\YtildeJ}{{\tilda Y}_{J}}

\def\beq{\begin{eqnarray}}
\def\eeq{\end{eqnarray}}
\def\beqn{\begin{eqnarray*}}
\def\eeqn{\end{eqnarray*}}

\def\hatt{\widehat}
\def\tilda{\widetilde}
\def\tr{{\rm t}}
\def\mse{{\rm mse}}
\def\d{{\rm d}}

\def\sumin{\sum_{i=1}^n}
\def\diag{{\rm diag}}
\def\bias{{\rm bias}}
\def\sqb{{\rm sqb}}
\def\full{{\rm full}}
\def\fic{{\rm fic}}
\def\wfic{{\rm wfic}}

\def\bfx{x}
\def\midd{\,|\,}

\author{Axel Gandy (Department of Mathematics, Imperial College London)\\
Nils Lid Hjort (Department of Mathematics, University of Oslo)}

\title{Focused Information Criteria for \\
 Semiparametric Linear Hazard Regression}

\begin{document}

\maketitle
\begin{abstract}
\noindent 
  The semiparametric linear hazard regression model introduced 
  by McKeague and Sasieni (1994) is an extension of the linear hazard
  regression model developed by Aalen (1980). Methods of model
  selection for this type of model are still underdeveloped. In the
  process of fitting a semiparametric linear hazard regression model
  one usually starts with a given set of covariates. For each
  covariate one has at least the following three choices: allow it to
  have time-varying effect; allow it to have constant effect over time; 
  or exclude it from the model. In this paper we discuss focused
  information criteria (FIC) to help with this choice. In the spirit
  of Claeskens and Hjort (2003), `focused' means that one is
  interested in one specific quantity, e.g.~the probability of
  survival of a patient with a certain set of covariates up to a given
  time. The FIC involves estimating the mean squared error of the 
  estimator of the quantity one is interested in, and the chosen model 
  is the one minimising this estimated mean squared error. 
  The focused model selection machinery is extended to allow
  for weighted versions, leading to a suitable wFIC method 
  that aims at finding models that lead to good estimates 
  of a given list of parameters, such as survival probabilities
  for a subset of patients 
  or for a specified region of covariate vectors. 
  In addition to developing model selection criteria, 
  methods associated with averaging across the best models 
  are also discussed. We illustrate these methods of model selection 
  in a real data situation.

  \smallskip\noindent
  {\em Key words and phrases:}
  Aalen's linear hazard regression model,
  focused model selection,
  model averaging,
  semiparametric,
  survival analysis
\end{abstract}


\section{Introduction}
\label{sec:intro}

We consider the linear additive risk regression model
for survival and event history analysis introduced 
by Aalen (1980), see e.g.~the discussion
Andersen et al.~(1993, Ch.~8) and 
in Martinussen and Scheike (2006, Ch.~5). 
Here the hazard rates (or intensity functions) 
for $n$ individuals are modelled as 
\beq
\label{eq:basic}
h_i(s) = x_i^\tr\alpha(s) 
       = x_{i,1}\alpha_1(s)+\cdots+x_{i,q}\alpha_q(s)
   \quad {\rm for\ }i=1,\ldots,n,
\eeq 
in terms of covariate information 
$x_i=(x_{i,1},\ldots,x_{i,q})^\tr$ for individual $i$.
Often, the first covariate is an intercept, i.e.~$x_{i,1}=1$.
The regression functions $\alpha_1(\cdot),\ldots,\alpha_q(\cdot)$
are the unknowns of the model, assumed to fulfill 
the basic requirement that $x^\tr\alpha(s)$ is
nonnegative for all $s\in[0,\tau]$ and for all 
vectors $x$ in the support of the distribution
of covariates; here $\tau$ is an upper limit on
the time scale involved, often chosen by convenience.

The Aalen model (\ref{eq:basic}) is most often 
applied in a nonparametric fashion, typically
leading to estimates and confidence bands 
for the cumulative regression functions 
$$A_j(t)=\int_0^t\alpha_j(s)\,\d s \quad {\rm for\ }j=1,\ldots,q $$
that avoid placing parametric restrictions 
on the $\alpha_j(\cdot)$ functions; 
see Section \ref{sec:examples} for an illustration. 
The model structure (\ref{eq:basic}) may however also 
be employed with some or all regression functions 
taken parametric. In particular, McKeague and Sasieni (1994) 
developed estimation methods for situations 
where some of the $\alpha_j(\cdot)$ are used 
nonparametrically (say for $j\in I$), 
and the others kept constant over time (say for $j\in J$);
here $I$ and $J$ are complementary subsets of $\{1,\ldots,q\}$.
The estimation strategy in question is reviewed in 
Section \ref{sec:semipar}. 

There is a wide literature touching both theoretical
and applied aspects of the Aalen model, 
but model selection issues, such as asking which
covariates need to be included in an analysis, 
appear not to have been studied until Hjort (2008),
where a certain focused information criterion (FIC)
was developed. Using that FIC enables one to 
decide which covariates ought to be respectively
included or excluded, from a longer list of
candidate covariates. The criterion applies 
for a given focus estimand, such as the survival probability
beyond a certain time point for a given type of patient,
and works via estimation of the mean squared error
involved for each of the $2^q$ candidate models
that may be considered. Weighted versions 
are also developed in Hjort (2008), 
the idea being
to select a model via a suitable wFIC score 
that produces precise estimates of a given list of estimands, 
such as the survival probability for a given subset 
of individuals or for a designated region 
of covariate vectors, or for a list of future time points 
for a given individual.

The focus of the present article is to extend these 
model selection criteria to the case 
where three different options exist for each covariate $x_j$: 
it may be included nonparametrically,
i.e.~as time-varying; 
it may be included parametrically, as constant over time;
and it may be excluded. 
Model selection therefore involves pinpointing 
three disjoint subsets of indices $\{1,\ldots,q\}$, 

-- $I$, those included in the model with time-varying effect; 

-- $J$, those included in the model with time-constant effect; and  

-- $K=\{1,\ldots,q\}-(I\cup J)$, those excluded from the model. 

\noindent 
Thus there is a total of $3^q$ candidate models to rank in order of
quality.  
Such a model selector leads to hazard rates of the form
$$h_i(s)=\sum_{j\in I}x_{i,j}\alpha_j(s)
   +\sum_{j\in J}x_{i,j}\alpha_j 
   \quad {\rm for\ }i=1,\ldots,n, $$
with corresponding cumulative hazard rates 
\beq
\label{eq:Hiofs}
H_i(s)=\sum_{j\in I}x_{i,j}A_j(s)
   +\sum_{j\in J}x_{i,j}\alpha_jt 
   \quad {\rm for\ }i=1,\ldots,n, 
\eeq 
to be used with the semiparametric estimation method 
mentioned above.  
The model selection criteria developed in the present article
are focused on estimating cumulative hazard
rates. 

The  estimates of the cumulative hazard rates  are often  used to produce
estimates of the survival probability of the form 
$$\hatt S_i(t)=\prod_{[0,t]}\{1-\d\hatt H_i(s)\}, $$
either for the existing individuals or for 
certain types of envisaged patients 
with covariance information of relevance. 
A model leading to precise estimation of cumulative 
hazard rates thus also leads to precise estimation
of survival probabilities. 

The general idea of FIC methods can be found in
Claeskens and Hjort (2003), 
with a wider discussion in Claeskens and Hjort (2008a).
A FIC for the multiplicative Cox model for survival data
(different in spirit and execution 
from the present additive Aalen model),
again with weighted wFIC versions, 
was developed and illustrated in Hjort and Claeskens (2006). 

In the present article, 
Section \ref{sec:semipar} provides the mathematical framework
inside which the basic FIC is being developed, 
giving for each focused question a ranked list 
of the $3^q$ candidate models. In particular, 
different focus parameters may lead to different rankings
and to differently selected top models. 
Weighted versions wFIC are developed in Section \ref{sec:weightFIC}, 
allowing some estimands to be judged more important
than others. The machinery is then illlustrated
in Section \ref{sec:examples}, applied to a survival data set
for 312 patients suffering from primary biliary cirrhosis,
a rare liver disease. Rather than sticking to only
the top model suggested by an application of the FIC or wFIC,
one may form weighted averages of estimates across
several top models. This model averaging topic is 
briefly discussed in Section \ref{sec:modelaveraging}. 
Finally, Section \ref{sec:concluding} offers 
some concluding remarks and pointers to further
research questions. 

\section{Theoretical development}
\label{sec:semipar}

We assume the classical setup for right censored survival data:
We observe $n$ individuals. For the $i$th individual, 
$T_i^0$ denotes the event time and  $C_i$ the censoring time.
We only observe
$T_i=\min(T_i^0,C_i)$ and $\delta_i=\ind\{T_i\leq C_i\}$,
along with a $q$-dimensional vector of covariates $x_i$.
The associated 0--1 counting processes of observed events 
are $N_i(t)=\ind\{T_i\leq t, \delta_i=1\}$ 
for $i=1,\ldots,n$. 

The additive model (\ref{eq:basic}) corresponds to 
the counting processes $N_i$ having intensity functions
$\lambda_i(s)=R_i(s)x_i^\tr\alpha(s)$,
where $R_i(s)=\ind\{T_i\ge s\}$ is the at-risk indicator.
In particular, the random processes $M_i$ with increments 
$\d M_i(s)=\d N_i(s)-\lambda_i(s)\,\d s$ 
become orthogonal zero-mean martingales,
see Andersen et al.~(1993, Ch.~II.4). 
This may be compactly represented as 
\beq
\label{eq:addmodel}
\lambda(s) = Y(s)\alpha(s),   
\eeq
where $\lambda(s)=(\lambda_1(s),\dots,\lambda_n(s))^\tr$ 
is the intensity of
$N=(N_1,\dots,N_n)^\tr$,
$Y(s)$ is the $n\times q$ matrix with 
rows $Y_i(s)=R_i(s)x_i^\tr$, and $\alpha(s)$ 
is the unknown $q$-dimensional parameter function of the model.
Note that $Y(s)$ is a previsible process 
in the parlance of martingale calculus; its paths
are left continuous and its values at time point $s$ 
are known just prior to $s$. 
The parameter space of the model is the set of 
such $\alpha(\cdot)$ functions for which
$x^\tr\alpha(s)$ is nonnegative 
for all $x$ in the support of the distribution
of covariate vectors; in particular, depending
on this distribution, some $\alpha_j(s)$ components
may be negative for some intervals of $s$ values.

At the beginning we want to estimate 
the cumulative hazard rate for an individual 
with covariate vector $x$, i.e.
$$ H(t,x)=x^\tr A(t)
   =x_1A_1(t)+\cdots+x_qA_q(t), $$
where $A(t) = \int_0^t \alpha(s)\,\dint s$
is the vector of $A_j(t)=\int_0^t\alpha_j(s)\,\dint s$. 
Instead of working with the full model we want to use a subset of the
covariates and/or restrict some covariates to have time-constant
effect in order to reduce the variance of the estimate,
as explained in the introduction. More precisely 
we wish to use the semiparametric model additive model by
McKeague and Sasieni (1994), 
where
$$\lambda(s) = \YI(s)\alpha_I(s) + \YJ(s)\alpha_J, $$
cf.~(\ref{eq:Hiofs}). Again, 
$I$ is the set of indices 
that have time-dependent effect (nonparametrically modelled)
and $J$ the set of indices of covariates 
that have time-independent effect (parametrically modelled).
Here $\YI(s)$ is the $n\times|I|$ submatrix of $Y(s)$ 
consisting of the columns associated with covariates from the set $I$
(and $|I|$ denoting the cardinality of $I$), 
and similarly for the $n\times|J|$ submatrix $\YJ(s)$.
As estimator for $H(t,x)$ we use 
\beq
\label{eq:hatHIJ}
\hatt H_{I,J}(t,\bfx)
   =\bfx_I^\tr\hatt{A}_I(t)+ 
    \bfx_J^\tr\hatt{\alpha}_Jt
   =\sum_{j\in I} x_j\hatt A_{I,j}(t)
    +\sum_{j\in J} x_j\hatt\alpha_{J,j}t, 
\eeq
where $\hatt{A}_I(t)$ and $\hatt{\alpha}_J$ 
are the least squares estimators from McKeague and Sasieni (1994), 
namely
\begin{align*}
\hatt{A}_I(t) 
  &=\int_0^t \{\YI(s)^\tr\YI(s)\}^{-1}\YI(s)^\tr
   \{\dint N(s)-\YJ(s) \hatt{\alpha}_J\,\dint s\},\\
\hatt{\alpha}_J
   &=\Bigl\{\int_0^\tau \YtildeJ(s)^\tr \YtildeJ(s)\,\dint s\Bigr\}^{-1}
   \int_0^\tau  \YtildeJ(s)^\tr\,\dint N(s),
\end{align*}
where $\YtildeJ(s)
=[\unitmatr_n -\YI(s)\{\YI(s)^\tr\YI(s)\}^{-1}\YI(s)^\tr]\,\YJ(s)$,
in terms of the identity matrix $\unitmatr_n$. 

We would wish to choose $I$ and $J$ such that the 
mean squared error of the (\ref{eq:hatHIJ}) estimator 
is minimised, i.e.~we want to minimise
\beq
\label{eq:msedecomposition}
\mse_n(I,J)
&=&\Expec \{\hatt H_{I,J}(t,\bfx)-H(t,\bfx)\}^2 \nonumber \\
&=&\Var\,\hatt H_{I,J}(t,\bfx) 
   +\{\bias\,\hatt H_{I,J}(t,\bfx)\}^2 \\
&=&v_n(I,J)+\sqb_n(I,J), \nonumber
\eeq
say, with $\sqb_n(I,J)$ being the square of 
the bias $\Expec\,\hatt H_{I,J}(t,\bfx)-H(t,\bfx)$. 
Our strategy is to estimate these two terms from data, 
and in the end select $(I,J)$ to minimise the 
consequent criterion. 

For the following derivations we assume that the additive model
(\ref{eq:addmodel}) is true.
We start out rewriting our estimator as
\beqn
\hatt H_{I,J}(t,\bfx)
=\int_0^\tau K(s)\,\dint N(s)
=\sumin \int_0^\tau K_i(s)\,\d N_i(s), 
\eeqn
where the $1\times n$ weight function may be expressed as 
\beqn
K(s)
   &=&\ind\{s\leq t\}\bfx_I^\tr\{\YI(s)^\tr\YI(s)\}^{-1}\YI(s)^\tr \\
& &+\left[t\bfx_J^\tr-\bfx_I^\tr\int_0^t 
   \{\YI(s)^\tr\YI(s)\}^{-1}\YI(s)^\tr \YJ(s)\,\dint s\right]
   \left\{\int_0^\tau  \YtildeJ(s)^\tr \YtildeJ(s)\,\dint  s\right\}^{-1}  
   \YtildeJ(s)^\tr
\eeqn
(omitting its dependence on $I$, $J$, $t$ in the notation). 
Consider now the decomposition
\beq
\label{eq:maindecomp}
\hatt H_{I,J}(t,\bfx)-H(t,\bfx)
= \left\{\int_0^\tau K(s)\,\dint\Lambda(s)
   -H(t,\bfx)\right\}
+\int_0^\tau K(s)\,\dint \{N(s) - \Lambda(s)\}.  
\eeq
For the second term, one can show using martingale calculus
and central limit theorems that 
$$\int_0^t K(s)
   \{\dint N(s) - \Lambda(s)\,\d s\} 
   =\int_0^t K(s)\,\d M(s) $$
is close in distribution to a zero mean Gaussian process
whose variance function $V_n(t)$ can be 
estimated by
$$\hatt v_n(I,J)
   =\int_0^\tau K(s)\,\diag\{\dint N(s)\}\,K(s)^\tr
   =\sumin \ind\{t_i\le t\}\delta_iK_i(t_i)^2. $$
More formally, $nV_n(t)$ converges to a well-defined
limit variance function, for which $n\hatt v_n(I,J)$ 
is a consistent estimator. 

The expected value  of the first term on the right hand 
side of (\ref{eq:maindecomp}) is  the bias. 
A natural bias estimator is  
$$\hatt b_n=\hatt H_{I,J}(t,\bfx)-\hatt H_\full(t,\bfx), $$
where 
$$\hatt H_\full(t,\bfx)=\hatt H_{\{1,\dots,q\},\emptyset}(t,\bfx)
  =\bfx^\tr\int_0^t\tilde K(s)\,\dint N(s) $$ 
is the full-model estimator of the cumulative hazard, with
$\tilde K(s)=\ind\{s \leq t\}
   \{Y(s)^\tr Y(s)\}^{-1}Y(s)^\tr$
the weight function involved in defining 
the estimator of $H(t,x)$ in the full model. 

We thus have an estimator $\hatt b_n$ being unbiased for the bias
$b_n$, and with a variance, say $\tau_n^2$, that may be estimated in
the manner above, replacing $K$ with $\tilde K$. This
leads to an asymptotically unbiased estimator of the squared bias:
\beqn \hatt\sqb_n(I,J)
&=&\hatt b_n^2-\hatt\tau_n^2 \\
&=&\{\hatt H_{I,J}(t,\bfx)-\hatt H_\full(t,\bfx)\}^2
-\int_0^\tau\{K(s) - \tilde K(s)\} \,\diag\{\dint
N(s)\}\, \{K(s) -\tilde K(s)\}^\tr.  \eeqn 
Going
back to (\ref{eq:msedecomposition}), then, this leads to the focused
information criterion
 \beq
\label{eq:fic1}
\fic_n(I,J)=\hatt v_n(I,J)+\max\{0,\hatt\sqb_n(I,J)\},
\eeq 
truncating if required negative estimates of the squared bias 
to zero. The FIC method, for the given covariate vector $\bfx$
and time point $t$, is to select the subsets $I$ and $J$
that minimise $\fic_n(I,J)$ across all candidates. 

\smallskip {\sl Remark A.}  In various applications it would make
sense to insist keeping some of the covariates inside the final model,
such as an intercept and perhaps `gender' and `treatment or control'.
This amounts to defining some covariates as `protected', limiting the
list of potential candidate models.  We may similarly insist on some
covariates being `protected and constant in time' while others may be
`protected and to be treated nonparametrically'.  This does not
disturb the machinery developed here, as it only amounts to a shorter
list of FIC values to compute and to rank.

\smallskip
{\sl Remark B.} 
Above the FIC is constructed in relation to 
the estimation of the full cumulative hazard 
rate from start of time up to some given $t$.
The machinery and formulae are easily modified
to the case of hazard cumulative increments 
of the type $H(t,\bfx)-H(t_0,\bfx)$,
of relevance when studying survival chances
beyond time $t$ for patients who have already
survived up to time $t_0$. 

\smallskip {\sl Remark C.}  The formulae given above are valid as long
as the required matrix inverses exist and the formulae produce
nonnegative and nondecreasing cumulative hazard estimates.  In
particular, one needs sufficiently many linearly independent covariate
vectors inside the risk sets.  The probability of this being satisfied
is often close to unity, and, in particular, this probability
converges exponentially quickly to 1 with increasing sample size, 
as long as there is a perhaps small 
but positive probability of having non-censored
life-lengths above the upper time window point $\tau$.

\section{Weighted FIC}
\label{sec:weightFIC}

The FIC strategy outlined above, leading to (\ref{eq:fic1})
as a model selection criterion, is for some application
overly focused, so to speak, as it specialises 
to the case of a given individual and a given time point.
More generally one may be interested in selecting a model
that works well for a certain stratum of individuals
(say women) and perhaps for a stretch of time points
(say age group 55 to 65). The present section develops
a weighted FIC approach for selecting a good model 
in such cases. 

Consider in general the loss function
\beqn
\ell_n(I,J)
&=&\sum_k w_n(t_k,\bfx_k)\{\hatt H_{I,J}(t_k,\bfx_k)-H(t_k,\bfx_k)\}^2 \\
&=&\int\{\hatt H_{I,J}(t,\bfx)-H(t,\bfx)\}^2
   \,\d W_n(t,\bfx),
\eeqn
with certain weights $w_n(t_k,\bfx_k)$ given 
to positions $(t_k,\bfx_k)$ in the space of $(t,\bfx)$.
The integral notation points to the possibility
of somewhat more general weighting schemes, 
e.g.~uniform weighting across a certain time window, etc.
We would wish to minimise the risk, i.e.~the expected loss $r_n(I,J)$. 
As long as these weights depend on covariate vectors
and time points, but not otherwise on data, we have
$$r_n(I,J)=\Expec\,\ell_n(I,J)
   =\int\mse_n(I,J,t,\bfx)\,\d W_n(t,\bfx), $$
with $\mse_n(I,J,t,\bfx)$ as in (\ref{eq:msedecomposition}),
but now explicitly taking $(t,\bfx)$ into the notation.
A natural estimate of this quantity emerges via
the separate variance and squared bias estimates, 
$$\hatt r_n(I,J)=\int\{\hatt v_n(I,J,t,\bfx)
   +\hatt\sqb_n(I,J,t,\bfx)\}\,\d W_n(t,\bfx). $$
Here we prefer to truncate the accumulated 
squared bias estimate to zero, if required,
as opposed to accumulating the individual truncated contributions.
The result is the weighted FIC selection criterion 
$$\wfic_n(I,J)=\int\hatt v_n(I,J,t,\bfx)\,\d W_n(t,\bfx)
   +\max\Bigl[0,\int\hatt\sqb_n(I,J,t,\bfx)\,\d W_n(t,\bfx)\Bigr]. $$

\smallskip
{\sl Remark D.}
The weight functions considered above are assumed
to be non-random; they may e.g.~depend on the 
covariate vectors of the data set, but not 
on the random life-lengths $t_i$, or on other
estimated parameters. 
The theory and formulae
may however be extended to allow also for 
random weight functions, under some conditions; 
cf.~Claeskens and Hjort (2008b). 

\smallskip
{\sl Remark E.}
 A particular choice of random  weights is given by choosing 
$W_n$ as the empirical distribution of the observed 
$(t,\bfx)$ values (with $t$ restricted to $[0,\tau]$):
$$\wfic_n(I,J)=n^{-1}\sumin\hatt v_n(I,J,\min(t_i,\tau),\bfx_i)
+\max\Bigl[0,n^{-1}\sumin \hatt\sqb_n(I,J,\min(t_i,\tau),\bfx_i) \Bigr]. $$ 
A problem of these weights is that they depend on the amount of
censoring.  
To avoid this, we suggest to replace for censored observations the
time $t_i$ essentially by the distribution of the full model conditional 
on survival up to time $t_i$.
In practical terms this amounts to generating $r$ samples
$t^\ast_{i,1},\dots,t^\ast_{i,r}$ from this conditional distribution 
and use the empirical distribution on $(\min(t_{i,j},\tau),x_i)$
over $i=1,\dots,n$ and $j=1,\dots,r$.
This weighting can be considered a good default choice. 

\section{Examples}
\label{sec:examples}

We consider the well-known PBC dataset from
Fleming and Harrington (1991), 
which consists of the follow-up of 312 randomised patients with
primary biliary cirrhosis, a rare autoimmune liver disease.  As
covariates we use an intercept, the treatment indicator, edema, sex,
age, bilirubin, albumin and prothrombin time.  The categorical
variables treatment indicator, presence of edema, and sex (male = 1) 
are all coded as 0 and 1. The covariates age, bilirubin, albumin and
prothrombin time have been centered.

\begin{figure}[tbp]
    \includegraphics[width=0.95\linewidth]{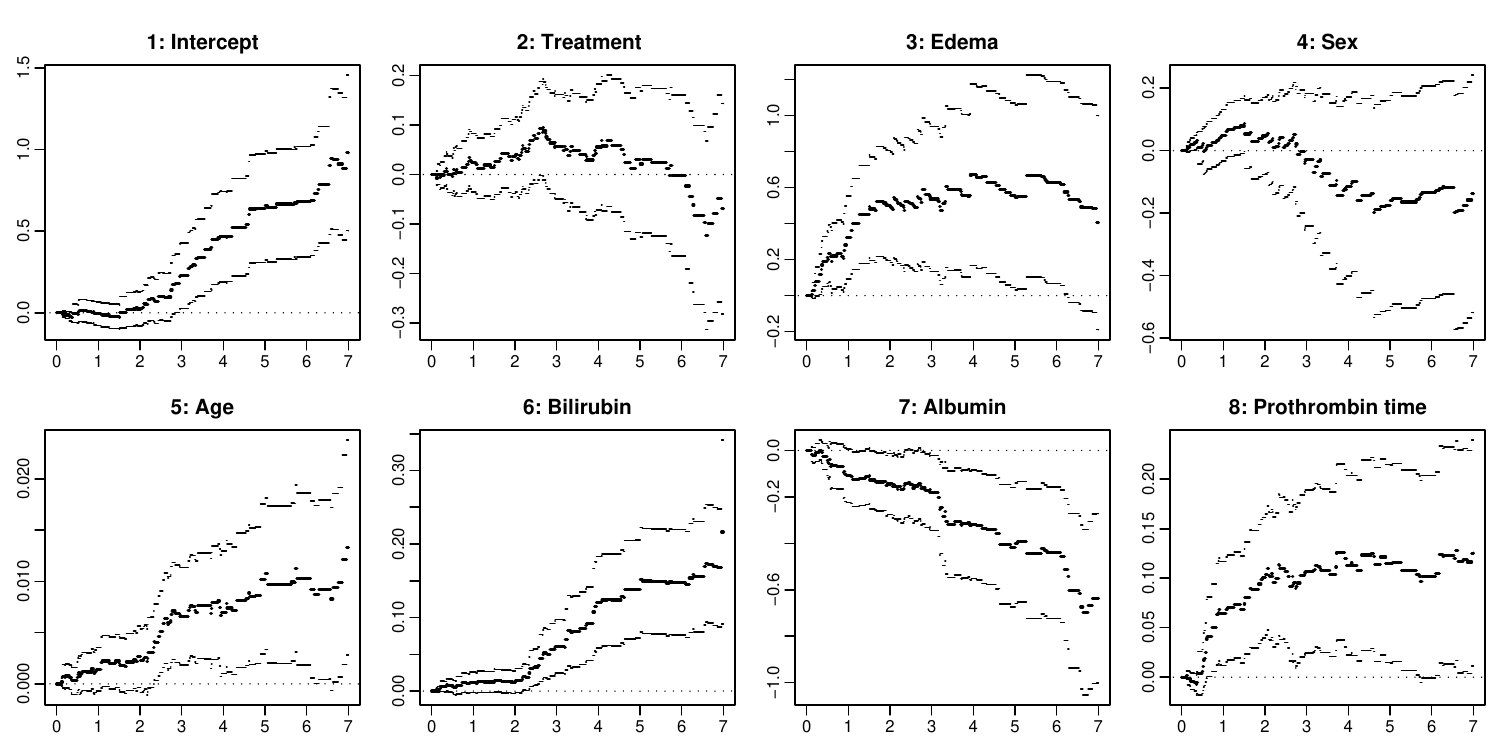}
    \caption{PBC data set: Estimates of cumulative regression 
      functions $A_j(t)$ based on the full additive model 
      with pointwise 95\% asymptotic confidence
      intervals. Time is in years.}
\label{fig:nonparpbc_complex}
\end{figure}

Figure \ref{fig:nonparpbc_complex} shows the fit of the full model.
We have included asymptotic pointwise 95\% confidence intervals.  In
these plots and in the following we use the endpoint $\tau=7$ years. 
After 7 years, 30\% of the patients are still at risk.

In the following we will protect the intercept and the treatment
indicator, i.e.~we only consider models that include both with
time-dependent effect.  This leaves us with six covariates, 
for each of which we have to decide between 
allowing a time-dependent effect, 
a time constant effect, and dropping the covariate from the model. 
Thus we consider $3^6=729$ potential models.

\subsection{FIC}

In this section we consider predicting the hazard rate 
$H$ at time $t=1$ for two different individuals.

The first individual is a treated 70 year old male with edema. His
bilirubin and prothrombin time are at the 75\% quantile and his
albumin is at the 25\% quantile among all individuals in the data set.
This makes him an individual with increased risk (see Figure
\ref{fig:nonparpbc_complex}).

The ordering of models according to the $\fic$ together with an FIC
plot can be seen in Figure \ref{fig:onecovarandtime_complex}.  An FIC
plot is a plot of the $\fic$ score against the prediction $\hat
H_{I,J}$ for all models under consideration.  The second and the third
column of the table contain the square roots of the variance and the
truncated estimated squared bias
$\hatt v^{1/2}$ and $(\hatt\sqb{}^+)^{1/2}=\max(\hatt\sqb,0)^{1/2}$,
where we have omitted the dependence on $n$, $I$ and $J$.
Of course, the sum of the squares of the second and third column is
precisely the $\fic$ in column 1, see (\ref{eq:fic1}),

The top model includes the four covariates intercept, treatment, sex and
age with time-dependent effect, and it includes edema with
time-constant effect. Note that all top models seem to be biased. In
fact, when comparing the estimates of the top model to the full model,
it seems that the top models give consistently lower estimates of
$\hatt H$ than the full model.

\begin{table}[tbp]
\centering
\begin{tabular}{r|r|r|l|l|r}
$\fic $&  $\sqrt{\hatt v}$&$\sqrt{\hatt\sqb{}^+}$&$I$&$J$&$\hatt H$\\\hline
  0.007 &   0.083 &   0.019 & 1,2,4,5 & 3&   0.389\\
  0.009 &   0.081 &   0.049 & 1,2,5 & 3&   0.376\\
  0.010 &   0.067 &   0.074 & 1,2,4 & 3,5&   0.350\\
  0.010 &   0.066 &   0.076 & 1,2 & 3,5&   0.349\\
  0.011 &   0.081 &   0.064 & 1,2,5 & 3,4&   0.368\\
  0.011 &   0.081 &   0.065 & 1,2,4,5 & 3,7&   0.368\\
  0.011 &   0.082 &   0.064 & 1,2,4,5,7 & 3&   0.369\\
  0.012 &   0.081 &   0.070 & 1,2,4,5 & 3,8&   0.365\\
  0.012 &   0.070 &   0.083 & 1,2,7 & 3,5&   0.348\\
  0.012 &   0.081 &   0.073 & 1,2,4,5,8 & 3&   0.363\\
\hline
  0.017 &   0.130 &   0.000 & 1,$\ldots$,8 & &   0.468\end{tabular}

\begin{minipage}{0.35\linewidth}
\includegraphics[width=\linewidth]{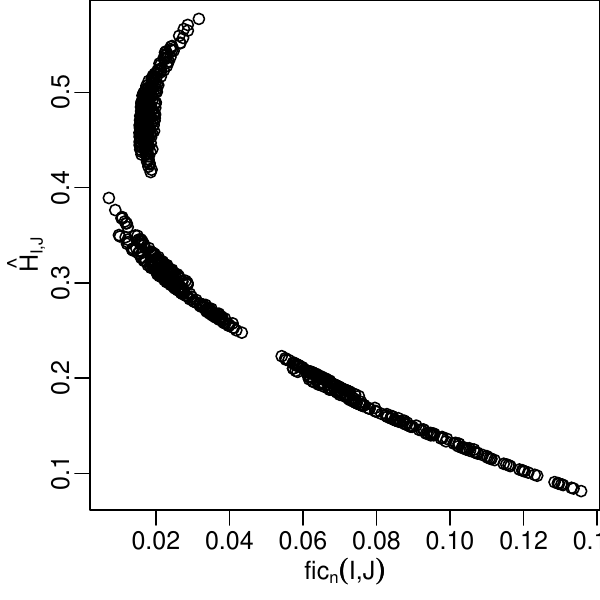}
\end{minipage}
\caption{ PBC data set: Ranking of models (last line: full model) 
  and corresponding FIC plot. Covariates are numbered 
  as in Figure \ref{fig:nonparpbc_complex}.
  We consider estimating the integrated hazard rate for a 70 year old
  male at time $t=1$ with higher-risk values of 
  bilirubin, albumin and prothrombin time.}
\label{fig:onecovarandtime_complexhighrisk}
\end{table}

The second individual we are considering is again a treated 70 year
old male with edema but now with average bilirubin, albumin and
prothrombin time.  The resulting ordering of models according to the
FIC is in Table \ref{fig:onecovarandtime_complex}.  The top model
includes the same covariates as before but now all of the
non-protected covariates are assumed to have time-constant effect.
Overall, the ordering of the models is somewhat different than for the
high-risk individual considered previously.

Note that some of the top models in Table
\ref{fig:onecovarandtime_complex} have an estimated squared bias equal
to $0$.  In fact, roughly 15\% of all the models under consideration
have an estimated squared bias equal to $0$.  This is not completely
surprising since we cut the estimate squared bias at $0$ and since we
are only trying to predict one quantity (and the biases due to the
omission of several covariates can cancel each other). 

\begin{table}[tbp]
\begin{tabular}{r|r|r|l|l|r}
$\fic $&  $\sqrt{\hatt v}$&$\sqrt{\hatt\sqb{}^+}$&$I$&$J$&$\hatt H$\\\hline
  0.004 &   0.066 &   0.000 & 1,2 & 3,4,5&   0.341\\
  0.004 &   0.066 &   0.000 & 1,2 & 3,5&   0.349\\
  0.004 &   0.067 &   0.000 & 1,2,4 & 3,5&   0.350\\
  0.006 &   0.064 &   0.038 & 1,2,4 & 3&   0.319\\
  0.006 &   0.077 &   0.004 & 1,2,4,5,6 & 3&   0.338\\
  0.006 &   0.064 &   0.045 & 1,2 & 3&   0.316\\
  0.006 &   0.080 &   0.000 & 1,2,4,5 & 3,7&   0.341\\
  0.006 &   0.081 &   0.000 & 1,2,4,5 & 3,8&   0.350\\
  0.007 &   0.081 &   0.000 & 1,2,5 & 3,4&   0.368\\
  0.007 &   0.081 &   0.000 & 1,2,5 & 3&   0.376\\
\hline
  0.016 &   0.128 &   0.000 & 1,$\ldots$,8 & &   0.419\end{tabular}

\begin{minipage}{0.35\linewidth}
\includegraphics[width=\linewidth]{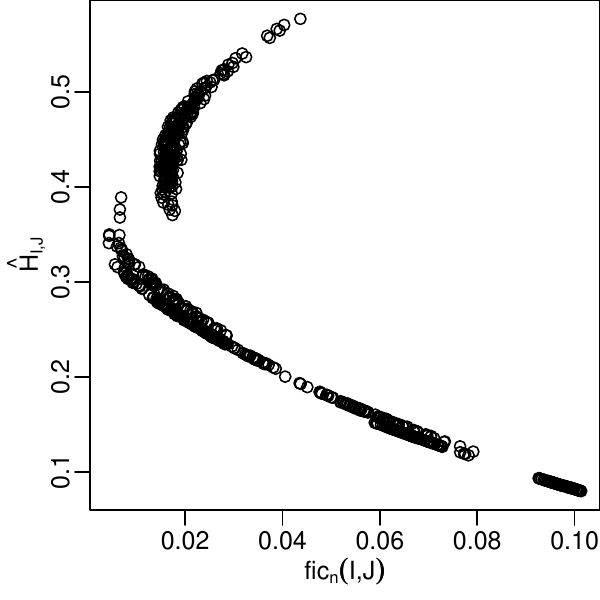}
\end{minipage}
\caption{Same setup as in Table \ref{fig:onecovarandtime_complexhighrisk},
  except that the patient has average values for bilirubin, albumin
  and prothrombin time.  }
\label{fig:onecovarandtime_complex}
\end{table}

\subsection{Weighted FIC}

In this subsection we consider using the weighted FIC ($\wfic_n$) as model
selection criterion. The weights $W_n$ are given by the empirical
distribution of 100 `virtual patients'. All of those 100 patients
are 50 year old male patients with edema. We call this part of the
covariate vector $x_{\rm fixed}$.  We wish to choose the remaining
covariates $x_{\rm rest}$ from the conditional distribution
$x_{\rm rest}\midd x_{\rm fixed}$.

We assume that the observed sample at hand is representative, so that
in particular the empirical covariance matrix for the $n$ individuals
is reasonable for the population at large and that the covariate
distribution is roughly multinormal.  Based on this assumption, we
sample $x_{\rm rest}$ for 100 virtual patients from the appropriate
conditional normal distribution.
For dichotomous covariates, such as the treatment indicator, the
assumption of normality is clearly not reasonable. To overcome this
we round the unconstrained realisation to the nearest possible value.


Tables \ref{tab:weightedFIC_maleed_t1} 
and \ref{tab:weightedFIC_maleed_t3} 
display results for predicting the integrated hazard 
for respectively $t=1$ and $t=3$ years. 
Note that we get different `winning' models. 
Also, 
there is no clear winner; 
several models have quite similar $\wfic_n$ scores.

Comparing the top models in Tables \ref{tab:weightedFIC_maleed_t1} 
and \ref{tab:weightedFIC_maleed_t3},
one can note that for $t=1$ more covariates have time-dependent effect 
than for $t=3$. In fact, for $t=1$ 
all top models have at least one time-dependent 
covariate beside the two protected covariates 1 and 2.

The column Mean($\hatt H$) (respectively SD($\hatt H$)) 
contain the mean (respectively standard deviation) 
of the 100 predictions from the model.

\begin{table}[tbp]
  \centering
  \begin{tabular}{r|r|r|l|l|r|r}
$\wfic$& $\sqrt{\int\hatt v\,\d W}$&$\sqrt{(\int\hatt\sqb\,\d W)^+}$&$I$&$J$&Mean($\hatt H$)&SD($\hatt H$)\\\hline
  0.015 &   0.099 &   0.073 & 1,2,4,8 & 3,5,6,7&   0.512&   0.138\\
  0.015 &   0.098 &   0.076 & 1,2,8 & 3,5,6,7&   0.510&   0.138\\
  0.017 &   0.101 &   0.081 & 1,2,4,6,8 & 3,5,7&   0.505&   0.120\\
  0.017 &   0.117 &   0.057 & 1,2,6,7,8 & 3,5&   0.533&   0.126\\
  0.017 &   0.117 &   0.058 & 1,2,4,6,7,8 & 3,5&   0.533&   0.127\\
  0.017 &   0.101 &   0.085 & 1,2,6,8 & 3,5,7&   0.502&   0.119\\
  0.018 &   0.097 &   0.092 & 1,2,8 & 3,4,5,6,7&   0.499&   0.138\\
  0.018 &   0.117 &   0.067 & 1,2,7,8 & 3,5,6&   0.532&   0.145\\
  0.018 &   0.117 &   0.069 & 1,2,4,7,8 & 3,5,6&   0.532&   0.144\\
  0.019 &   0.116 &   0.073 & 1,2,6,7,8 & 3,4,5&   0.524&   0.126\\
\hline
  0.029 &   0.170 &   0.000 & 1,$\ldots$,8 & &   0.634&   0.104\end{tabular}

  \caption{Ranking of models according to the weighted FIC 
    for 50 year old male patients with edema
    when estimating the integrated hazard rate at $t=1$.  
    Top 10 models. Last line is the full model.}
  \label{tab:weightedFIC_maleed_t1}
\end{table}

\begin{table}[tbp]
  \centering
  \begin{tabular}{r|r|r|l|l|r|r}
$\wfic$& $\sqrt{\int\hatt v\,\d W}$&$\sqrt{(\int\hatt\sqb\,\d W)^+}$&$I$&$J$&Mean($\hatt H$)&SD($\hatt H$)\\\hline
  0.054 &   0.232 &   0.000 & 1,2 & 3,5,6,7&   1.409&   0.330\\
  0.054 &   0.233 &   0.000 & 1,2 & 3,5,6,7,8&   1.408&   0.344\\
  0.055 &   0.231 &   0.041 & 1,2,4 & 3,5,6,7,8&   1.380&   0.341\\
  0.055 &   0.234 &   0.020 & 1,2,6 & 3,5,7&   1.418&   0.316\\
  0.055 &   0.235 &   0.000 & 1,2,6 & 3,5,7,8&   1.418&   0.331\\
  0.055 &   0.231 &   0.047 & 1,2 & 3,4,5,6,7,8&   1.380&   0.343\\
  0.056 &   0.236 &   0.000 & 1,2,8 & 3,4,5,6,7&   1.410&   0.355\\
  0.056 &   0.233 &   0.043 & 1,2,6 & 3,4,5,7,8&   1.393&   0.330\\
  0.056 &   0.237 &   0.000 & 1,2,4,8 & 3,5,6,7&   1.412&   0.352\\
  0.056 &   0.233 &   0.043 & 1,2,4,6 & 3,5,7,8&   1.391&   0.328\\
\hline
  0.080 &   0.282 &   0.000 & 1,$\ldots$,8 & &   1.514&   0.314\end{tabular}

  \caption{Weighted FIC scores for the integrated hazard rate at $t=3$.
    Otherwise the setup is  as in Table \ref{tab:weightedFIC_maleed_t1}.
}
  \label{tab:weightedFIC_maleed_t3}
\end{table}

\begin{figure}[tbp]
\centering
\includegraphics[width=0.45\linewidth]{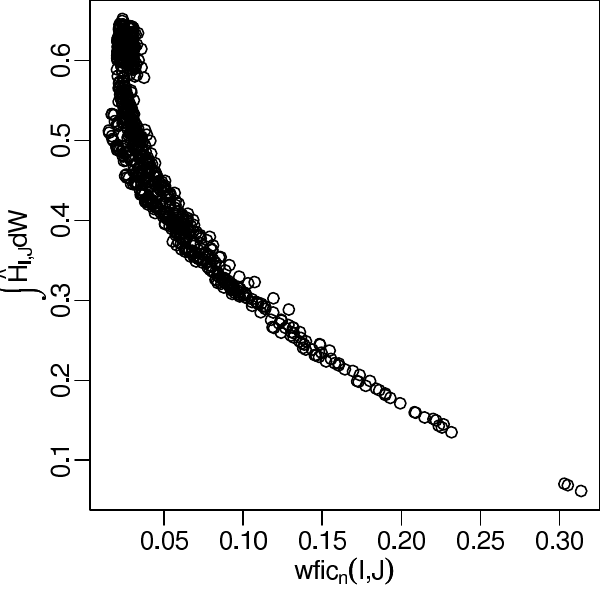}
\includegraphics[width=0.45\linewidth]{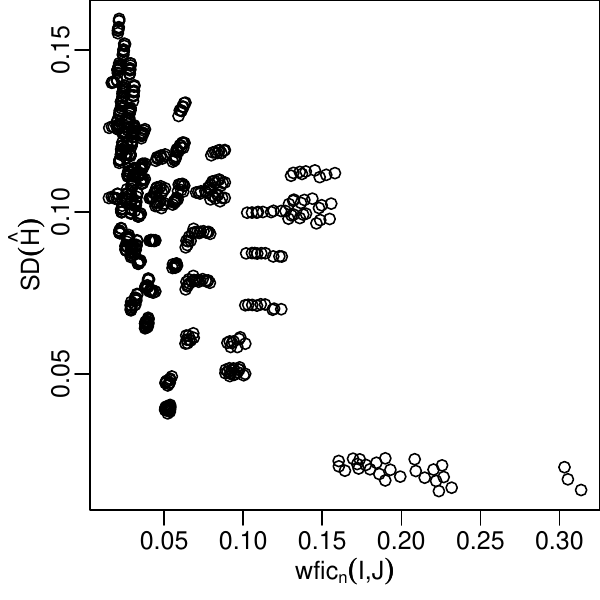}
\caption{Plots corresponding to Table \ref{tab:weightedFIC_maleed_t1}.
 wFIC plot and plot of $\wfic_n$ against the standard deviation 
 of the resulting estimates.}
\label{fig:wFICplot_t1}
\end{figure}

Figure \ref{fig:wFICplot_t1} contains the canonical analogue of FIC
plots for the weighted FIC based on the same setup 
as in Table \ref{tab:weightedFIC_maleed_t1}.
It shows a plot of the $\wfic_n$ against the mean
prediction. Furthermore, it contains a second plot that shows
the spread of the predictions. 
The plots further underline that there is no clear winner -- increasing
the appeal of model averaging techniques that will be discussed in
Section \ref{sec:modelaveraging}.

\subsection{Empirical weighted FIC}

Table \ref{tab:weightedFIC_empirical} contains the FIC of models based
on the empirical distribution as described in Remark E at the  end of Section
\ref{sec:weightFIC} with $r=1$. Thus we use the weights $(t_i^\ast,x_i)$, 
where for uncensored observations $t_i^\ast=\min(t_i,\tau)$ and 
for censored observation $t_i^\ast$ is the minimum of $\tau$ 
and a random variable generated from the conditional distribution 
given $t_i$, assuming independent censoring under the full model.

All of the top model include covariate 3, presence of edema, as
time-dependent covariate.  This is consistent with the fit of the full
model in Figure \ref{fig:nonparpbc_complex}, where covariate 3 seems
to have a strong effect initially that disappears after roughly 1.5
years.

\begin{table}[tbp]
  \centering
  \begin{tabular}{r|r|r|l|l|r|r}
$\wfic$& $\sqrt{\int\hatt v\,\d W}$&$\sqrt{(\int\hatt\sqb\,\d W)^+}$&$I$&$J$&Mean($\hatt H$)&SD($\hatt H$)\\\hline
  0.024 &   0.138 &   0.066 & 1,2,3 & 5,6,7,8&   0.390&   0.460\\
  0.025 &   0.140 &   0.071 & 1,2,3,5 & 6,7,8&   0.388&   0.459\\
  0.025 &   0.142 &   0.068 & 1,2,3,7 & 5,6,8&   0.394&   0.467\\
  0.025 &   0.139 &   0.075 & 1,2,3,8 & 5,6,7&   0.394&   0.451\\
  0.026 &   0.146 &   0.066 & 1,2,3 & 4,5,6,7,8&   0.391&   0.464\\
  0.026 &   0.144 &   0.071 & 1,2,3,5,7 & 6,8&   0.392&   0.466\\
  0.026 &   0.142 &   0.077 & 1,2,3,7,8 & 5,6&   0.397&   0.459\\
  0.026 &   0.141 &   0.079 & 1,2,3,5,8 & 6,7&   0.391&   0.450\\
  0.027 &   0.146 &   0.073 & 1,2,3,8 & 4,5,6,7&   0.394&   0.456\\
  0.027 &   0.147 &   0.072 & 1,2,3,4 & 5,6,7,8&   0.392&   0.461\\
\hline
  0.031 &   0.176 &   0.000 & 1,$\ldots$,8 & &   0.402&   0.517\end{tabular}

  \caption{Weighted fits. Weights are the 
  empirical distribution (adjusted for censoring -- see text) 
  of all individuals. Top 10 models.
  }
  \label{tab:weightedFIC_empirical}
\end{table}


\section{Model averaging}
\label{sec:modelaveraging}

As the examples worked through above demonstrate, 
there are often several models that score almost equally
well on the FIC and wFIC scales. It is then tempting
to construct suitable model average estimators, 
say of the form 
$$\tilda H(t,\bfx)=\sum_{m=1}^M c(m)\hatt H_{I_m,J_m}(t,\bfx), $$
where $(I_m,J_m)$ for $m=1,\ldots,M$ represent the index sets 
for the $M$ most promising models. 
The weights $c(m)$ should sum to one, 
and could e.g.~simply be $1/M$, but it is natural to 
make models that score well have higher weights. 
See Hjort and Claeskens (2003) 
for a general discussion
of such model average estimators, where weighting
schemes of the form 
$$c(m)={\exp\{-\lambda\,\fic_n(I_m,J_m)\}
   \over \sum_{m'=1}^M \exp\{-\lambda\,\fic_n(I_{m'},J_{m'})\}} $$
are advocated and seen to work well in terms
of the final risk functions. 
Here $\lambda$ is an algorithmic parameter,
with higher values giving more relative weight
to the best models. Inference based on such
estimators is non-standard due to the complicated
probability distributions involved, 
but may employ bootstrapping from the estimated 
largest model; see Remarks A and B of Section \ref{sec:concluding}.


\section{Discussion and concluding remarks}
\label{sec:concluding}

This final section offers some supplementary 
concluding remarks, some of which may invite
further research efforts. 

\smallskip
{\sl A. Behaviour of the final estimator.}
The properties of each submodel based estimator 
$\hatt H_{I,J}(t,\bfx)$ are quite well understood
from earlier investigations in the literature; 
in particular, each such is approximately normally
distributed for large sample sizes, with 
a variance that may be consistently estimated. 
The problem is harder when it comes to the 
finally selected estimator 
$$\tilda H(t,\bfx)=\hatt H_{\hatt I,\hatt J}(t,\bfx), $$
in that the selected subsets involved are also random.
Its distribution is approximately that of 
a non-linear mixture of correlated normals 
(a linear combination of a multinormal vector, 
where the weights depend on this vector
and hence are random themselves), and is, in particular,
not approximately normal itself. 
Assessing and estimating its distribution 
and e.g.~bias and standard deviation are accordingly 
challenging tasks. 
Similar non-trivial problems are associated
with the more general model average estimators
considered in Section \ref{sec:modelaveraging}. 
Methods of Hjort and Claeskens (203) and 
Claeskens and Hjort (2008a, Ch.~7)
may be used to attack these problems. For supplementing 
estimating with confidence intervals of 
test inference statements, bootstrapping may 
be used, which we describe next. 

\smallskip
{\sl B. Bootstrapping.}
We noted above that $\tilda H(t,\bfx)$ has a complicated
distribution. Write 
$$G_n(z,A_1,\ldots,A_q)=\Prob\{\tilda H(t,\bfx)-H(t,\bfx)\le z
   \midd A_1,\ldots,A_q\} $$
for the distribution of the error $\tilda H(t,\bfx)-H(t,\bfx)$,
exhibiting in the notation the dependence 
upon the unknown regressor functions $A_1,\ldots,A_q$. 
We now advocate estimating this distribution 
by plugging in the consistent 
estimators $\hatt A_{1,\full},\ldots,\hatt A_{q,\full}$ from the full model,
i.e.~the nonparametric Aalen estimators that
correspond to $I=\{1,\ldots,q\}$ and $J=\emptyset$. 

This amounts to bootstrapping from the largest estimated model,
and the result is 
$$\hatt G_n(z)=G_n(z,\hatt A_{1,\full},\ldots,\hatt A_{q,\full})
   ={\Prob}^*\{\tilda H^*(t,\bfx)-\hatt H_\full(t,\bfx)\le z
   \midd \hatt A_{1,\full},\ldots,\hatt A_{q,\full}\} , $$
with ${\Prob}^*$ indicating the probability mechanism
in which data and in particular $\hatt H_\full$ are kept fixed
and where $\tilda H^*(t,\bfx)$ is computed exactly 
as $\tilda H(t,\bfx)$, complete with estimated subsets
$\hatt I^*$ and $\hatt J^*$, but via simulated data sets
of the form $(T_i^*,\delta_i^*,{\bfx}_i)$ for $i=1,\ldots,n$.
Here $T_i^*=\min(T_i^{0,*},C_i^*)$, with 
$T_i^{0,*}$ drawn from the estimated distribution 
$\hatt F_i(t)=1-\prod_{[0,t]}\{1-\d\hatt H_i(s)\}$,
and $C_i^*$ is either the known censoring time, 
if appropriate, or drawn from the Kaplan--Meier estimated
censoring distribution. Note that ${\bfx}_i$ is kept equal
to its known value when producing $T_i^*$ and 
$\delta_i^*=\ind\{T_i^{0,*}\le C_i^*\}$. 
Via such bootstrapping one may now estimate the 
mean squared error 
$\Expec\{\hatt H_{\hatt I,\hatt J}(t,x)-H(t,x)\}^2$ 
and also set confidence intervals, as follows. 
Let $c_n<d_n$ be the $\alpha$ and $1-\alpha$ 
quantile points from $\hatt G_n$ above, 
i.e.~from the bootstrap distribution 
of $\tilda H^*-\hatt H_\full$. Then 
$[\tilda H-d_n,\tilda H-c_n]$ is an approximate
$1-2\alpha$ confidence interval for $H$. 
It is assumed here that the largest Aalen model 
(the one keeping each $A_j$ in the model as
a time-varying function) is adequate. 

\smallskip
{\sl C. Implied tests.}
The FIC and wFIC methods yield ranked lists of 
candidate models, and hence implied answers 
to questions of the type 
`should covariate $x_j$ be moved from $J$ to $I$?'.
There is accordingly and implied test statistic,
testing the hypothesis that $\alpha_j(\cdot)$
is constant over time vs.~the alternative
that it is not. Similar questions involve
the possibility of moving $x_j$ from $K$ to $I$,
or from $K$ to $J$, where the model selection
methods may be translated to statistics
for testing the hypothesis $\alpha_j=0$
vs.~two different types of alternatives
(nonzero and constant, and nonzero and time-varying).
These implied tests do have significance levels
determined by various aspects of the model,
and are not necessarily small -- just as 
there are tests and associated significance levels 
implied by the use of the Akaike Information Criterion AIC 
in parametric models.

\smallskip
{\sl D. Estimators using estimated optimal weights.}
The methodology we have developed in this article 
has been based on the estimators of McKeague and Sasieni (1994), 
cf.~(\ref{eq:hatHIJ}). These may be likened 
to `direct martingale-based least squares estimators'.
It is however possible to employ some more elaborate
weighted martingale-based least squares estimators,
where the weights in question are estimated versions
of the theoretically optimal weights, which 
by necessity also need to involve nonparametric
estimation of the regression functions 
$\alpha_1,\ldots,\alpha_q$ of (\ref{eq:basic}) 
themselves (as opposed to their cumulatives).
Our model selection methodology may be extended
to cover also these more refined estimators. 
The machinery is however rather more technically demanding,
and it is not clear to what extent precision 
would really improve, even if theorems may be
given that secure such improvement for very 
large sample sizes, cf.~Huffer and McKeague (1991). 

\smallskip {\sl E. FIC relatives.}  We constructed our FIC criterion
(\ref{eq:fic1}) via a truncation to zero of the underlying estimate of
squared bias. It is however possible to keep the squared bias estimate
on board untouched, which would then lead to a moderately different
FIC relative, say $\fic_n^*(I,J)$ instead of $\fic_n(I,J)$. What is
best among these two (and yet further) alternatives may be studied
from different perspectives -- one may study the probabilities of
hitting the underlying optimal submodel, or the precision of the final
estimator $\hatt H_{\hatt I,\hatt J}$.  Some analysis and comparisons
of this sort are carried out in Claeskens and Hjort (2003), and 
indicate that there can be no clear overall winner; the two resulting
risk functions would merely win in different parts of the parameter
space.


\subsection*{Acknowledgments}

The authors would like to express their gratitude for being invited to
take part in the stimulating environment of the Centre for Advanced
Studies at the Academy of Sciences and Letters in Oslo, where Odd
Aalen and \O rnulf Borgan organised a programme on `Statistical
Analysis of Complex Event History Data' for the academic year
2005--2006.



\bigskip
\parindent0pt 
\parskip3pt 

Aalen, O.O. (1980).
A model for nonparametric regression analysis of counting processes.
In {\em Mathematical Statistics and Probability Theory -- 
Proceedings, Sixth International Conference, Wisla (Poland)}, 
volume~2 of {\em Lecture Notes in Statistics}, 
pp.~1--25. Springer-Verlag, New York.

Andersen, P., Borgan, \O., Gill, R., and Keiding, N. (1993).
{\em Statistical Models Based on Counting Processes}.
Springer Verlag, New York.

Claeskens, G.~and Hjort, N.L. (2003).
{The focused information criterion.}
{\em Journal of the American Statistical Association}
  {\bf 98}, 901--916.

Claeskens, G.~and Hjort, N.L. (2008a).
{\em Model Selection and Model Averaging.}
Cambridge University Press, Cambridge.

Claeskens, G.~and Hjort, N.L. (2008b).
{Minimising average risk in regression models.}
{\em Econometric Theory} {\bf 24}, 493--527.

Fleming, T. and Harrington, D. (1991).
{\em Counting Processes and Survival Analysis}.
Wiley, New York.

Hjort, N.L.~and Claeskens, G. (2003).
{Frequentist Model Average Estimators.}
{\em Journal of the American Statistical Association}
  {\bf 98}, 879--900.

Hjort, N.L.~and Claeskens, G. (2006).
Focused information criteria and model averaging 
for the {C}ox hazard regression model.
{\em Journal of the American Statistical Association}
  {\bf 101}, 1449--1464.

Hjort, N.L. (2008).
Focused information criteria for the linear hazard regression model.
In {\em Statistical Models and Methods 
for Biomedical and Technical Systems}, 
pp.~487--502. Birkh\"auser, Boston.

Huffer, F.W.~and McKeague, I.W. (1991).
Weighted least squares estimation for Aalen's additive risk model.
{\sl Journal of the American Statistical Association} {\bf 86},
114--129. 

Martinussen, T.~and Scheike, T. (2006).
{\em Dynamic Regression Models for Survival Data}.
Springer-Verlag, Heidelberg. 

McKeague, I.~W.~and Sasieni, P.~D. (1994).
A partly parametric additive risk model.
{\em Biometrika} {\bf 81}, 501--514.


\end{document}